\documentclass[12pt]{iopart}

\expandafter\let\csname equation*\endcsname\relax
\expandafter\let\csname endequation*\endcsname\relax
\usepackage{amsmath}

\usepackage{amssymb}
\usepackage[amssymb]{SIunits}
\usepackage{graphicx}
\usepackage{xspace}
\usepackage{bm}
\usepackage[colorlinks,urlcolor=blue]{hyperref}
\usepackage[all]{hypcap}
\newcommand*{\doi}[1]{\href{http://dx.doi.org/\detokenize{#1}}{\detokenize{#1}}}

\begin{document}

\title{FIREFLY: heat load and particle exhaust approximations for rapid evaluation of divertor designs}

\author{H. Frerichs${}^1$, D. Boeyaert${}^1$, Y. Feng${}^2$, D. Reiter${}^3$}

\address{${}^1$ Department of Nuclear Engineering \& Engineering Physics, University of Wisconsin - Madison, WI, USA}
\address{${}^2$ Max-Planck-Institut f\"ur Plasmaphysik, Greifswald, Germany}
\address{${}^3$ Institute for Laser and Plasma Physics, Heinrich-Heine-University, D-40225 D\"usseldorf, Germany}

\ead{hfrerichs@wisc.edu}

\begin{abstract}
The divertor in a magnetic confinement fusion reactor is an essential component for power dissipation and particle removal.
The FIREFLY package for rapid evaluation of divertor designs is presented as an extension of the FLARE code for field line reconstruction from a flux tube mesh.
First, divertor loads are approximated with a simplified heat transport model.
Neutralized particles are then sampled from the resulting load distribution, and the EIRENE code is used to track molecules and atoms in a plasma background while accounting for dissociation, charge exchange and ionization.
Particles are removed on pumping surfaces in order to estimate the exhaust efficiency for a given divertor geometry.
Optimization of the divertor geometry for more efficient particle exhaust is explored by using W7-X as an example, and the sensitivity to model parameters for the plasma background in the proxy calculations is evaluated.
\end{abstract}

\vspace{2pc}

\def\vec#1{\ensuremath{{\bf #1}}\xspace}
\def\Poincare{Poincar\'e\xspace}
\def\qt{\ensuremath{q_{t}}\xspace}
\def\qproxy{\ensuremath{h_{t}}\xspace}
\def\PSOL{\ensuremath{P_{\mathrm{SOL}}}\xspace}
\def\nsepx{\ensuremath{n_{\mathrm{sepx}}}\xspace}
\def\bstep{\ensuremath{\Delta_{\parallel}}\xspace}
\def\xstep{\ensuremath{\Delta_{\perp}}\xspace}
\def\pload{\mbox{\footnotesize \ensuremath{\mathcal{P}}}\xspace}
\def\ploss{\ensuremath{p_{\mathrm{loss}}}\xspace}
\def\nbg{\ensuremath{n_{0}}\xspace}
\def\Tbg{\ensuremath{T_{0}}\xspace}
\def\ncore{\ensuremath{n_{\mathrm{core}}}\xspace}
\def\Tcore{\ensuremath{T_{\mathrm{core}}}\xspace}
\def\nedge{\ensuremath{n_{\mathrm{edge}}}\xspace}
\def\Tedge{\ensuremath{T_{\mathrm{edge}}}\xspace}
\def\gpump{\ensuremath{g_{\mathrm{pump}}}\xspace}
\def\gcore{\ensuremath{g_{\mathrm{core}}}\xspace}
\def\gfast{\ensuremath{g_{\mathrm{fast}}}\xspace}

\def\el{\mathrm{e}\xspace}
\def\Datm{\mathrm{D}\xspace}
\def\Dmol{\mathrm{D}_{2}\xspace}
\def\Dmli{\mathrm{D}_{2}^{+}\xspace}
\def\Dion{\mathrm{D}^{+}\xspace}

\def\epump{\ensuremath{\varepsilon_{\mathrm{pump}}}\xspace}
\def\wgap{\ensuremath{w_{\mathrm{gap}}}\xspace}
\def\qpeak{\ensuremath{\max \qproxy}\xspace}
\def\qpumplimit{\ensuremath{q_t|_{\mathrm{pump}}^{(\mathrm{limit})}}\xspace}

\def\vmax{\ensuremath{\vec{v}_{\textnormal{max}}}\xspace}

\section{Introduction}

The divertor is an essential component for power and particle control in magnetic confinement fusion plasmas \cite{Loarte2007, Krieger2025} - not just in tokamaks but also in stellarators \cite{Koenig2002}.
Power exhaust from the main plasma into the scrape-off layer (SOL) can result in extreme heat loads, and divertor targets must be designed to withstand those and protect the main chamber walls.
In addition to that, divertor baffles are needed to support the removal of neutralized particles before they can return to the main plasma.
Potential advantages of stellarator configurations over tokamaks are an improved stability and the intrinsic design for continuous operation, but this comes at the expense of significantly increased complexity.
Particular challenges resulting from this complexity are the toroidal localization of heat loads and inefficient particle removal due to insufficient compression of neutral particles in the divertor volume.

Driving the plasma in front of the divertor targets into a {\it detached} state \cite{Matthews1995, Krasheninnikov2017, Stangeby2018} can significantly reduce heat loads.
This is achieved by gas puffing into the SOL and divertor volume, and it is likely to include seeded impurities (N, Ne, Ar, ...) for power dissipation in fusion reactors and upcoming experiments.
In fact, radiation from seeded impurities is an integral part of the ITER divertor design \cite{Pacher2015, Pitts2019}.
High-fidelity modeling of detachment (i.e. self-consistent plasma - neutrals gas - impurity interactions) is indispensable for successful divertor design, and for stellarators this requires 3D models such as the EMC3-EIRENE code suite \cite{Feng2004} with its recent upgrades for recombination processes \cite{Frerichs2021, Feng2025}.
However, this is computationally expensive and it is suitable only for a small number of potential divertor configurations for design verification.
Low-fidelity models (i.e. with simplified physics such as EMC3-Lite \cite{Feng2022} or HEAT \cite{Looby2022}), on the other hand, may be sufficient to capture essential trends with respect to the divertor geometry.
Consequently, they can be utilized for rapid evaluation of potential divertor designs during optimization for selected figures of merit.
For example, minimizing the peak heat load in the absence of power dissipation can be useful because it implies minimizing the need for impurity seeding and its detrimental side effects.

The FIREFLY package for optimization of the divertor geometry is presented.
It is based on the FLARE code \cite{Frerichs2024a} for construction of a magnetic flux tube mesh.
Such a mesh enables other codes (such as EMC3) to reconstruct magnetic field lines \cite{Feng2005} which provides significant speedup over numerical integration.
Recently, a semi-automatic mesh generation method for unstructured layouts \cite{Frerichs2025} has been implemented into FLARE in support of field line diffusion and heat load proxy calculations.
%
%
These calculations are summarized in section \ref{sec:divertor_load_proxies} where the impact of model parameters is evaluated by using W7-X as an example.
The novel aspects of FIREFLY address the matter of particle exhaust.
Approximation of the exhaust efficiency is introduced and evaluated in section \ref{sec:particle_exhaust_proxy}.
In essence, neutral particles are generated based on the approximated divertor load distribution, and they are then followed until they are either ionized or removed at a pumping surface.
Optimization of the divertor geometry is discussed in section \ref{sec:divertor_shape_optimization}.
First, we explore simple adjustments of the divertor geometry in parameter scans.
Then, we integrate the proxy calculations into a multivariate optimization procedure to maximize particle removal while avoiding heat loads to the potential pump gap.


\section{Divertor load proxy} \label{sec:divertor_load_proxies}

Since plasma transport along magnetic field lines is several orders of magnitude faster than in cross-field direction, divertor loads are closely linked to the magnetic geometry.
In poloidal divertor tokamaks, the strike locations of the magnetic separatrix mark the positions where particle and heat loads are found.
The island divertor in stellarators extends this concept to more complex non-axisymmetric configurations.
However, it can be more challenging to clearly identify the magnetic separatrix as it is prone to chaos (often referred to as stochasticity) \cite{Tomita1978}.
In any case, cross-field transport will smooth out divertor loads around the strike locations.
Thus, a first approximation of divertor loads can be found by introducing artificial cross-field diffusion to magnetic field lines as described below for reference.
However, a better approximation may be achieved based on a simplified heat transport model which is discussed in section \ref{sec:heat_load_proxy}.
Nevertheless, either approximation can be used as input for the particle exhaust proxy (section \ref{sec:particle_exhaust_proxy}).
Both approximations leverage field line reconstruction from a flux tube mesh which covers the volume between good flux surfaces (somewhere near the last closed flux surface) and plasma facing components (PFCs).

\subsection{Strike point density}

The purpose of introducing artificial cross-field diffusion to magnetic field lines is to mimic plasma transport.
Field lines are launched randomly on the inner boundary of the flux tube mesh.
This corresponds to energy flux \PSOL from the core plasma which is a boundary condition for EMC3-EIRENE simulations.
After each step \bstep along a field line, a displacement (i.e. cross-field jump)

\begin{equation}
\xstep \, = \, \sqrt{2 \, \mathcal{D} \, \bstep} \, \bm{\xi}_\perp \label{eq:xstep_D}
\end{equation}

\noindent is introduced based on a random vector $\bm{\xi}_\perp$ with 2-D normal distribution.
For convenience, $\bm{\xi}_\perp$ is taken in the r-z plane. 
%
%
%
%
Diffusive field lines are followed until they intersect PFCs.
The resulting strike point density \pload as proxy for divertor loads is normalized to the total field line count, i.e. $\int\!d\!A \, \pload \, = \, 1$.

The field line diffusion coefficient $\mathcal{D}$ is a free parameter, but it can be associated with a particle or heat diffusion coefficient $\chi$ based on a characteristic velocity $u$ for advection along field lines.
Let $\tau \, = \, \bstep / u$ be a time step, then it follows from (\ref{eq:xstep_D}) that $\chi \, = \, \mathcal{D} \, u$ which results in the same cross-field step (\ref{eq:xstep_chi}) that we introduce later for heat conduction.
For example, $\mathcal{D} \, = \, 10^{-5} \, \meter^2 \, \meter^{-1}$ corresponds to $\chi \, = \, 1 \, \meter^2 \, \second^{-1}$ for H ions at thermal velocity at $100 \, \electronvolt$, and a given value for $\chi$ implies larger field line diffusion at colder temperatures.
Results have been verified against the equivalent procedure in FLARE based on numerical integration instead of the significantly faster field line reconstruction \cite{Frerichs2025}.

\subsection{Simplified heat transport} \label{sec:heat_load_proxy}

For an improved approximation of heat loads, we turn to a simplified heat transport model.
The model is the same as in EMC3-Lite \cite{Feng2022} except that the implementation here is based on an unstructured mesh.
This makes it more suitable for strongly shaped stellarator configurations.
The model assumes that heat conduction dominates over convection and that losses from excitation and ionization of neutral particles and impurity ions are negligible.
This yields the following heat transport equation:

\begin{equation}
\nabla \cdot \left(-\kappa \, \nabla_\parallel T \, - \, \chi \, n \, \nabla_\perp T \right) \, = \, 0 \label{eq:simplified_heat_transport}
\end{equation}

\noindent with equal temperature $T \, = \, T_e \, = \, T_i$ and density $n \, = \, n_e \, = \, n_i$ for electrons and ions.
The total anomalous cross-field heat conductivity for electrons and ions $\chi \, = \, \chi_e \, + \, \chi_i$ is assumed to be spatially constant and is a free parameter for the model.
Classical heat conductivity along field lines is much smaller for ions than for electrons such that $\kappa \, \approx \, \kappa_e(T) \, = \, \kappa_{e0} \, T^{5/2}$ where $\kappa_{e0}$ is a physical constant.
The model is further simplified by setting $\kappa \, = \, \kappa_e(\Tbg)$ to be constant, along with $n = \nbg$.
Even though this assumption is strictly only applicable to low density plasmas where no significant temperature gradients exist along field lines, it should be noted that we are here interested mainly in the heat loads \qt on the divertor targets.
Thus, we should interpret \nbg and \Tbg as the {\it heat transport averaged} plasma density and temperature that yield the same heat loads as the full non-linear model - even at higher density with significantly different upstream and downstream temperatures.

We can solve (\ref{eq:simplified_heat_transport}) by leveraging the relation to the transition probability of a stochastic process and apply a Monte Carlo algorithm - essentially the same procedure as in EMC3 \cite{Feng2000}.
But now that (\ref{eq:simplified_heat_transport}) is linear in $T$, it can be solved without iterations.
Also, the results are proportional to \PSOL and we can introduce the heat load proxy $\qproxy \, = \, \qt / \PSOL$ which has the same units as the strike point density \pload. 
We divide (\ref{eq:simplified_heat_transport}) by $n$ such that $\kappa / n$ is a diffusivity, and arrive at the following steps for transport along and across field lines:
%
%
\begin{eqnarray}
\bstep  & = & \delta_\parallel \, \xi_\parallel \quad \mathrm{with} \quad \delta_\parallel \, = \, \sqrt{2 \, \kappa/n \, \tau} \label{eq:bstep} \\
\xstep  & = & \sqrt{2 \, \chi \, \tau} \, \bm{\xi}_\perp \label{eq:xstep_chi}
\end{eqnarray}

\noindent where $\xi_\parallel$ is a random number with normal distribution and $\bm{\xi}_\perp$ is again a random vector in the r-z plane.
Hence, the simplified heat transport model is very similar to the strike point density calculation, except that the step along field lines \bstep is also of diffusive nature here.
The other difference is the boundary condition on PFCs.
The Bohm condition

\begin{equation}
\qt \, = \, \gamma \, n \, c_{st} \, T_t
\end{equation}

\noindent with sound speed $c_{st} \, = \, c_s(T_t)$ and $\gamma = \gamma_e + \gamma_i \, \approx \, 7$ can be expressed as a loss probability for the incident Monte Carlo particles:

\begin{equation}
\ploss \, = \, \sqrt{\frac{\pi}{2}} \, \frac{\delta_\parallel}{\lambda_{T \parallel}}
\quad \mathrm{with} \quad
\lambda_{T \parallel} \, = \, \frac{\kappa}{n \, c_s \, \gamma} \label{eq:ploss}
\end{equation}

\noindent This is accurate to the first order of $\delta_\parallel / \lambda_q$ which requires that the time step $\tau$ is sufficiently small \cite{Feng2022}.

\begin{figure}
\begin{center}
\includegraphics[width=160mm]{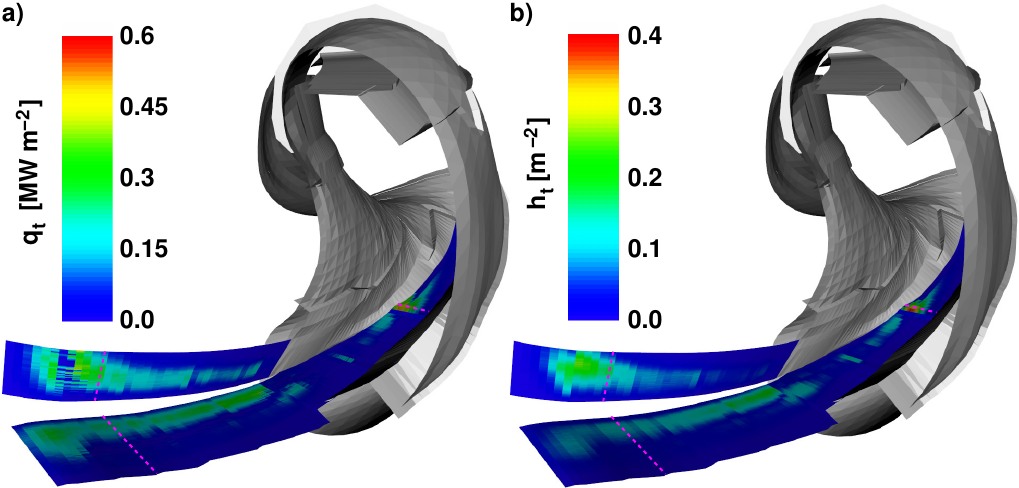}
\caption{(a) Heat loads \qt on the divertor targets in W7-X computed by EMC3-EIRENE for $\PSOL \, = \, 3 \, \mega\watt$ with $20 \, \%$ impurity radiation \cite{Boeyaert2024}. An additional $25 \, \%$ of \PSOL is lost to neutrals. Contributions from surface recombination that is usually included in the EMC3-EIRENE post-processing have been switched off here for direct comparison with \qproxy. (b) Heat load proxy \qproxy computed by FIREFLY with $\nbg = 2 \, \cdot 10^{19} \, \meter^{-3}$ and $\Tbg \, = \, 16 \, \electronvolt$.}
\label{fig:ht_best_fit}
\end{center}
\end{figure}

Monte Carlo particles (which represent energy flux from the core) are initialized randomly along the inner mesh boundary and are followed until they are eventually deposited on the PFCs.
Small \ploss implies that they tend to bounce back several times before they are deposited.
Whether or not the resulting \qproxy is a good approximation of the expected heat loads depends on the choice of the free parameters \nbg, \Tbg and $\chi$.
From (\ref{eq:bstep}) we expect equivalent heat transport along field lines for $\Tbg^{5/2} \, \sim \, \nbg$, but the boundary condition (\ref{eq:ploss}) may lead to some deviation due to the weak temperature dependence of $c_s$.
In the following we will evaluate choices for \nbg and \Tbg based on an EMC3-EIRENE simulation for W7-X in standard divertor configuration.
This particular simulation has been conducted for $\PSOL \, = \, 3 \, \mega\watt$ and $\chi_e \, = \, \chi_i \, = \, 3 \, \meter^2 \, \second^{-1}$ with $20 \, \%$ radiative power losses from intrinsic carbon.
The density at (just inside) the separatrix is set to $\nsepx = 1.7 \, \cdot \, 10^{19} \, \meter^{-3}$ and anomalous cross-field diffusion is set to $D \, = \, 1 \, \meter^2 \, \second^{-1}$.
This is the lowest density within a series of simulations that have been conducted and evaluated earlier \cite{Boeyaert2024}.
The resulting heat load on the divertor targets is shown in figure \ref{fig:ht_best_fit} (a) for reference.

In order to evaluate the simplified heat transport model, we first have to account for the total power losses in the EMC3-EIRENE simulation.
On top of the $20 \, \%$ losses from intrinsic carbon, and additional $25 \, \%$ are lost to neutral particles that are present from recycling on the divertor targets.
Thus, the net power in the simulation is $\PSOL^\ast \, \approx \, 1.64 \, \mega\watt$.
We then define

\begin{equation}
\chi^2 \, = \, \sum\left(\qt^{\mathrm{(EMC3-EIRENE)}} / \PSOL^\ast \, - \, \qproxy\right)^2 \label{eq:chisq}
\end{equation}

\noindent as the difference between the EMC3-EIRENE simulation and the simplified heat transport model.
The sum in (\ref{eq:chisq}) is taken over all cells of the surface mesh on each of the divertor targets.

\begin{figure}
\begin{center}
\includegraphics[width=160mm]{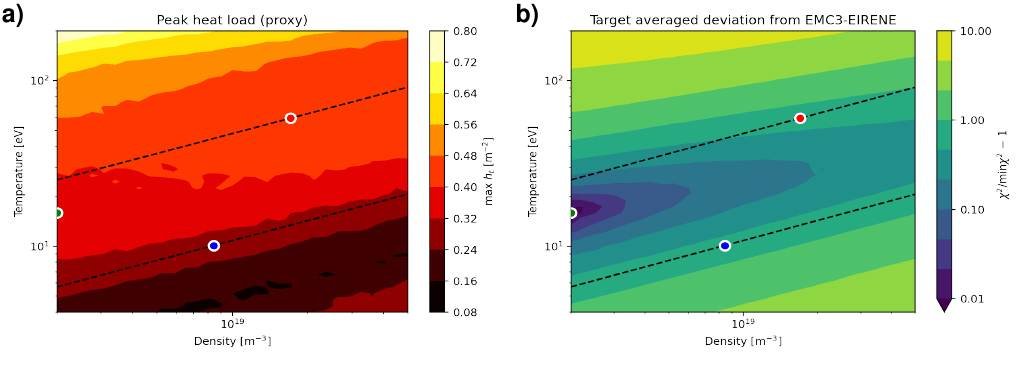}
\caption{(a) Proxy for peak heat load $\max h_t$ and (b) target averaged deviation from EMC3-EIRENE.
Colored dots mark the conditions in the EMC3-EIRENE simulation at the separatrix (red) and averaged on the divertor targets (blue).
The deviation is normalized to the best fit $\min\chi^2$ within this scan (green dot).
Black dashed lines indicate $\kappa/n$ = const.}
\label{fig:hload_parameters}
\end{center}
\end{figure}

We set $\chi \, = \, 6 \, \meter^2 \, \second^{-1}$ in the simplified heat transport model to match the cross-field heat conduction in the EMC3-EIRENE reference simulation.
Results from a 2D scan over \nbg and \Tbg are shown in figure \ref{fig:hload_parameters}, which indicates that \nbg and \Tbg should be selected somewhere between separatrix and target values.
Within this range, the peak heat load proxy varies from $\max h_t \, = \, 0.29 \, \meter^{-2}$ to $0.43 \, \meter^{-2}$.
The black dashed lines show that $\kappa / n$ = const yield largely similar results, but with a trend towards better fits at lower density.
Remarkably, the best fit is achieved at the lowest density in the scan, which is even lower than the averaged density at the divertor targets in the EMC3-EIRENE reference simulation.
This is perhaps not so much a consequence of the simplified heat conduction along field lines (spatially constant $\kappa$), but rather a consequence of the missing heat convection terms in the simplified model and the spatial dependence of power losses in the EMC3-EIRENE simulation.
Also, it may be necessary to introduce a separate set of plasma parameters for the boundary condition (\ref{eq:ploss}).

The heat load proxy \qproxy for the best fitting \nbg and \Tbg is shown in figure \ref{fig:ht_best_fit} (b) which shows overall good agreement with the EMC3-EIRENE simulation.
One concern may be that heat loads are missing in some areas on the left side of the vertical target in figure \ref{fig:ht_best_fit} (a), which may lead to artificial contributions to $\chi^2$ that vary with \nbg and \Tbg.
This is where the strike direction of field lines change and is most likely due to insufficient mesh resolution in this EMC3-EIRENE simulation and the discrete approximation of divertor targets therein.
Nevertheless, the nature of the results is not affect when $\chi^2$ is instead evaluated only along selected profiles at $\varphi = \pm 13 \, \deg$ as highlighted in figure \ref{fig:ht_best_fit}.


\section{Particle exhaust proxy} \label{sec:particle_exhaust_proxy}

While the divertor targets (and plasma scenario) must be designed to withstand heat loads, optimizing for \qt alone is not sufficient for a successful divertor design.
Another requirement for the divertor is that that neutral particles - recycled plasma or sputtered impurities - must be removed before they can enter the core plasma.
One possible complication is charge exchange with plasma ions which can produce energetic neutrals, and those can impact the first wall or other unprotected surfaces that are not designed for plasma exposure.
Thus, we need additional modeling to support divertor design.

\begin{figure}
\begin{center}
\begin{minipage}{0.35\textwidth}
\includegraphics[width=50mm]{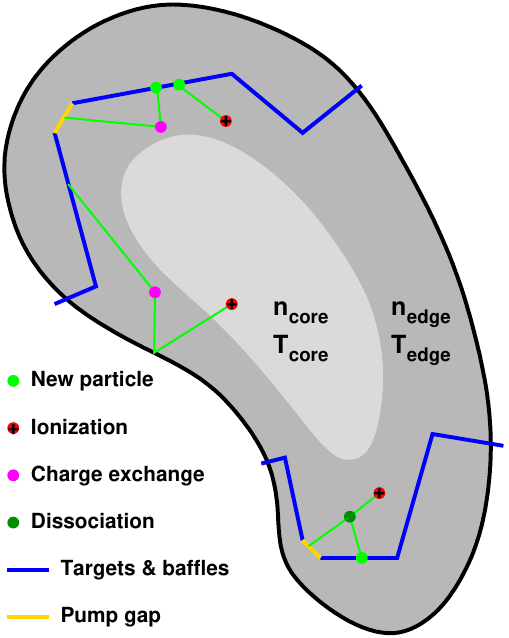}
\end{minipage}\hfill
\begin{minipage}{0.65\textwidth}
\footnotesize
\begin{center}\begin{tabular}{llll}
Reaction & & & Description \\
\hline
$\Datm \, + \, \el$     & $\rightarrow$ &   $\Dion \, + \, 2 \el$                  & Ionization \\
$\Datm \, + \, \Dion$   & $\rightarrow$ &   $\Dion \, + \, \Datm$                  & Charge exchange \\
$\Dmol \, + \, \el$     & $\rightarrow$ &   $2 \Datm \, + \, \el$                  & Dissociation \\
$\Dmol \, + \, \el$     & $\rightarrow$ &   $\Dmli \, + \, 2 \el$                  & Non-dissociative ionization \\
$\Dmol \, + \, \el$     & $\rightarrow$ &   $\Datm \, + \, \Dion \, + \, 2 \el$    & Dissociative ionization \\
$\Dmli \, + \, \el$     & $\rightarrow$ &   $2 \Dion \, + \, 2 \el$                & Dissociative ionization \\
$\Dmli \, + \, \el$     & $\rightarrow$ &   $\Datm \, + \, \Dion \, + \, \el$      & Dissociative excitation \\
$\Dmli \, + \, \el$     & $\rightarrow$ &   $2 \Datm$                              & Dissociative recombination \\
\end{tabular}\end{center}
\end{minipage}
\caption{(a) A few example trajectories of neutral particles and their reactions. (b) List of reactions that are included in the simplified particle exhaust calculation.}
\label{fig:reactions}
\end{center}
\end{figure}

Motivated by the simplified heat transport model from the previous section, we now introduce a simplified model for particle exhaust (figure \ref{fig:reactions}).
For now, we focus on plasma particles which are neutralized on PFCs, and assume that their initial distribution is proportional to \qproxy or \pload.
Neutral particles are followed until they are either ionized or reach a pumping surface.
The simulation domain for neutral particles is divided into two reservoirs: a core domain and an edge/SOL domain.
The plasma density and temperature within each reservoir is assumed to be spatially constant.
The edge/SOL domain is where we applied the simplified heat transport model, and one may be inclined to set $\nedge = \nbg$ and $\Tedge = \Tbg$.
However, one should interpret \nbg and \Tbg as the net plasma conditions for heat transport while \nedge and \Tedge are the net plasma conditions for neutral particles.
Because of the opposite source locations (core vs. PFCs) and different transport characteristics, those parameters are not necessarily the same.

Self consistent recycling is not included as this would require iterations between particle and energy transport, i.e. results here are linear with respect to the total recycling flux $\Gamma$.
Computed quantities are the ionization source in the core \gcore (as a proxy for neutral compression in the divertor volume), the pumped particle flux \gpump, and the incident flux of energetic particles \gfast on each PFC, all relative to $\Gamma$.
The threshold for energetic particles is set to $10 \, \electronvolt$ here.
In order to explore the usefulness of this approach, we use the EIRENE code \cite{Reiter2005} with specialized interface routines.
Within EIRENE, the TRIM database is used for the surface reflection model \cite{Eckstein1991}, and the HYDHEL \cite{Janev1987} and AMJUEL \cite{Greenland1996} databases are used for reaction rates.
Atoms and molecules can undergo a number of reactions as summarized in figure \ref{fig:reactions} (b).
This is a reduced set of table 1 in reference \cite{Kotov2008} and does not include el.-ion recombination, ion conversion ($\Dmol \, + \, \Dion \, \rightarrow \, \Dmli \, + \, \Datm$) and elastic collisions.
Nevertheless, this reduced set is often used in EMC3-EIRENE simulations for W7-X (including the one used here for reference).

\begin{figure}
\begin{center}
\includegraphics[width=160mm]{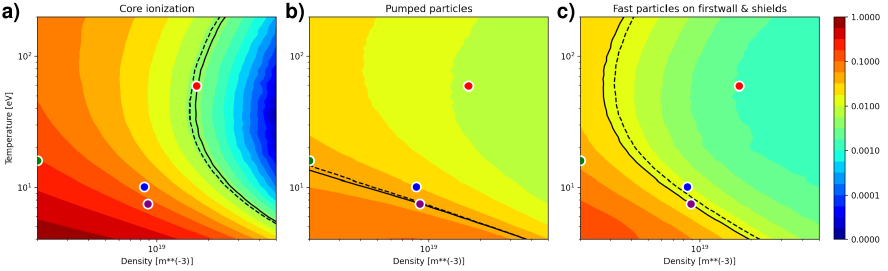}
\caption{Computed quantities in the simplified particle exhaust calculation based on \qproxy from figure \ref{fig:ht_best_fit} (b): (a) ionization source in the core \gcore, (b) pumped particle flux \gpump, and (c) incident flux of energetic particles \gfast on the first wall and shields in W7-X.
The background plasma in the core is set to $\ncore = 5 \, \cdot \, 10^{19} \, \meter^{-3}$ and $\Tcore = 200 \, \electronvolt$.
Solid lines show the corresponding value in the EMC3-EIRENE reference calculation. Dashed lines show the same contour lines for another parameter scan with $\ncore = 10^{20} \, \meter^{-3}$ and $\Tcore = 1 \, \kilo\electronvolt$.
Colored dots mark the plasma conditions in the EMC3-EIRENE simulation at the separatrix (red) and averaged on the divertor targets (blue), as well as the plasma conditions with the best fitting \qproxy result (green) and where both \gpump and \gfast match their reference values (purple).
}
\label{fig:pexhaust_parameters}
\end{center}
\end{figure}

In the following we will explore the impact of the choice of the background plasma.
For the purpose of evaluating the divertor geometry with respect to its particle exhaust efficiency, we replace the gap between the horizontal and vertical targets with a pumping surface and set the removal probability to $\epump \, = \, 100 \, \%$.
While this does not account for particles returning from the sub-divertor volume in realistic configurations, our focus is on the ability to remove particles from the divertor volume.
Furthermore, for the impact of fast particles, we will focus on the first wall and shields in W7-X.
Figure \ref{fig:pexhaust_parameters} shows the dependence of \gcore, \gpump and \gfast on \nedge and \Tedge where \qproxy from figure \ref{fig:ht_best_fit} with $\nbg \, = \, 2 \, \cdot \, 10^{19} \, \meter^{-3}$ and $\Tbg \, = \, 16 \, \electronvolt$ is used as input.
Contour lines for matching conditions in the EMC3-EIRENE reference simulation are shown in black.
As is turns out, the dependence on \ncore and \Tcore is rather weak, which is not unreasonable as long as \Tcore is high enough for ionization to dominate over charge exchange.
We find that \gpump in the proxy calculation is compatible with that in the EMC3-EIRENE reference simulation when setting $\nedge = \nbg$ and $\Tedge = \Tbg$, but the corresponding \gcore and \gfast values are different.
As mentioned above, we should not expect that the same net plasma conditions are applicable to heat conduction and neutral particles transport.
And indeed, the contour lines for equivalent results in figure \ref{fig:pexhaust_parameters} are quite different from the ones in figure \ref{fig:hload_parameters}.
No particular $(\nedge, \Tedge)$ combination can reproduce all three parameters from the EMC3-EIRENE simulation at the same time, but at least both \gpump and \gfast are found to match their reference values under conditions (purple) that are close to the target averaged ones (blue).
On the other hand, upstream conditions (red) produce a better result for \gcore, and thus the best choice for (\nedge, \Tedge) depends on the given priorities for optimization.
In any case, what needs to be confirmed is how well \gcore, \gpump and \gfast capture trends with respect to changes in the target and baffle geometry relative to the magnetic configuration.
This is beyond the present scope, but additional analysis for different configurations (high-iota, low-iota, ...) would be beneficial as follow-up.


\section{Divertor shape optimization} \label{sec:divertor_shape_optimization}

The simplified models presented in sections \ref{sec:divertor_load_proxies} and \ref{sec:particle_exhaust_proxy} are intended for rapid evaluation of divertor designs for the purpose of shape optimization.
Thus, in order to make good use of these capabilities, we also need a suitable representation of the divertor geometry.
This must include enough fidelity to capture the essential properties (such as shallow field line incident), while avoiding too many degrees of freedom.
We turn again to the W7-X divertor as an example and consider the geometry used in EMC3-EIRENE simulations.
Both horizontal and vertical targets are given as straight plates in the r-z plane with $0.5 \, \deg$ resolution in toroidal direction.
This results in a total number of 540 shape coefficients for the geometry in figure \ref{fig:ht_best_fit} (not counting additional baffles and heat shields, which were included in the simulations there but are neglected in this section).

Rather than tuning those shape coefficients individually, we introduce a parametrization of the geometry based on figure \ref{fig:shape_parameters} (a).
For this we consider the cross-section of the divertor to be a function of the toroidal angle, and we will use cubic splines for a fit to the actual geometry.
For the present analysis we use 7 equidistant knots between $-18 \, \deg$ and $28 \, \deg$.
Even though this may miss some of the details of the current W7-X divertor configuration, it can be seen in figure \ref{fig:shape_parameters} (b) that the heat load distribution is captured fairly well.
Moreover, this smoothes over some of the features of the tail end which reduces the peak load proxy from $0.38 \, \meter^{-2}$ to $0.24 \, \meter^{-2}$.
Additional knots could be introduced to fine tune the fit as a starting point for optimization of heat load spreading.
This would be a major application that we leave for a future project.
Here, we rather focus on improving the efficiency of particle removal while avoiding heat loads to potential pump gaps.
First, we explore simple adjustments of the current geometry in parameter scans.
Then, we integrate the proxy calculations into a multivariate optimization procedure.

\subsection{Exploration of basic adjustments}

\begin{figure}
\begin{center}
\includegraphics[width=160mm]{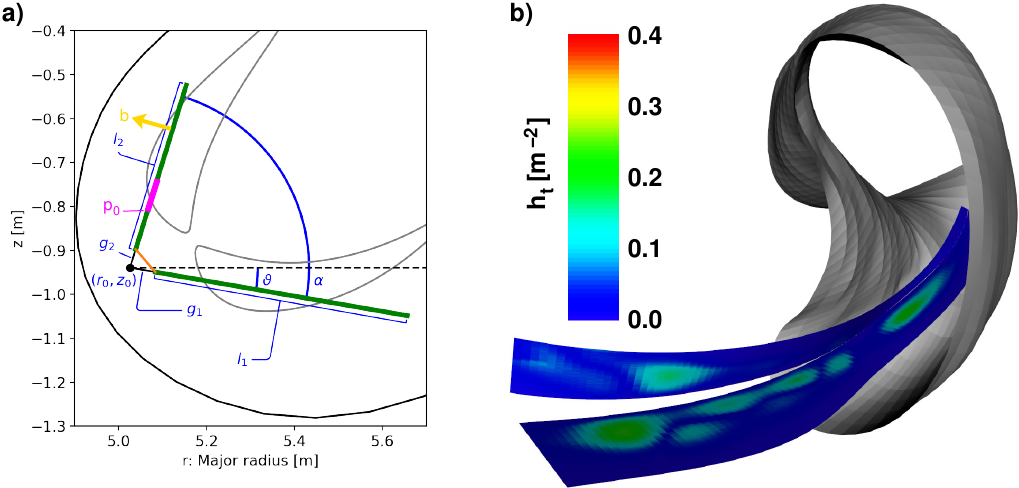}
\caption{(a) Shape parameters (blue) for the horizontal and vertical targets (green) and pump gap (orange) of the island divertor in W7-X. These parameters are functions of the toroidal angle. Additional shape adjustments for the vertical target position (yellow) and pump gap position (magenta) are highlighted. (b) Heat load proxy for the parameterized divertor shape fitted to the current W7-X geometry using cubic splines with 7 equidistant knots. The same model parameters (\nbg, \Tbg, $\chi$) as in figure \ref{fig:ht_best_fit} (b) are applied here.}
\label{fig:shape_parameters}
\end{center}
\end{figure}

The vertical target intercepts another island chain near $\varphi = 6-12 \, \deg$ where it receives a peak heat load proxy of $0.17 \, \meter^{-2}$.
In the following we will evaluate the impact of pushing back the vertical target such that it rather becomes a baffle.
Specifically, we will push back each cross-section of the vertical target by

\begin{equation}
b(\varphi) \, = \, b_0 \, \exp\left(\frac{-(\varphi - \varphi_0)^2}{2 \, \sigma^2}\right) \label{eq:offset}
\end{equation}

\begin{figure}
\begin{center}
\includegraphics[width=160mm]{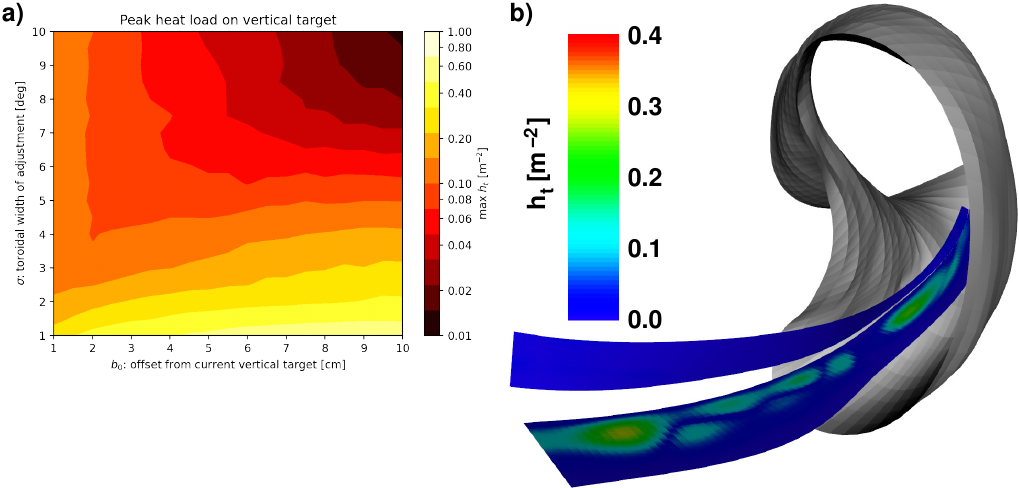}
\caption{(a) Parameter scan for the peak heat load on the vertical target with shape adjustments $(b_0, \sigma)$. (b) Resulting heat load proxy for $b_0 = 10 \, \centi\meter$ and $\sigma \, = \, 10 \, \deg$.}
\label{fig:vertical_target_adjustment}
\end{center}
\end{figure}

\noindent in the normal direction within each r-z plane as highlighted in yellow in figure \ref{fig:shape_parameters}.
Large $\sigma$ implies that the plates is pushed back evenly, while small $\sigma$ implies a local adjustment around $\varphi_0 \, = \, -9.5 \, \deg$.
Figure \ref{fig:vertical_target_adjustment} the resulting peak heat loads on the vertical target.
For local adjustments with $\sigma = 1 \, \deg$, we find that this results in localized hot spots with increasing peak loads with increasing $b_0$.
This is related to the local field line incident angle which becomes less grazing the more curvature is introduced around $\varphi_0$.
At larger sigma ($\sigma > 5 \, \deg$), on the other, we find that the peak load is reduced with increasing $b_0$.
The heat load on the first wall remains below $1 \, \%$ in all cases.
Figure \ref{fig:vertical_target_adjustment} (b) shows the resulting heat load distribution with minimal heat loads on the vertical target.
It can be seen that a side effect of minimizing the peak head load on the vertical target is an increase of the peak heat load on the horizontal target.
Nevertheless, this may be acceptable if it allows to improve particle exhaust by moving the pump entrance upwards where it may capture more particles.

\begin{figure}
\begin{center}
\includegraphics[width=160mm]{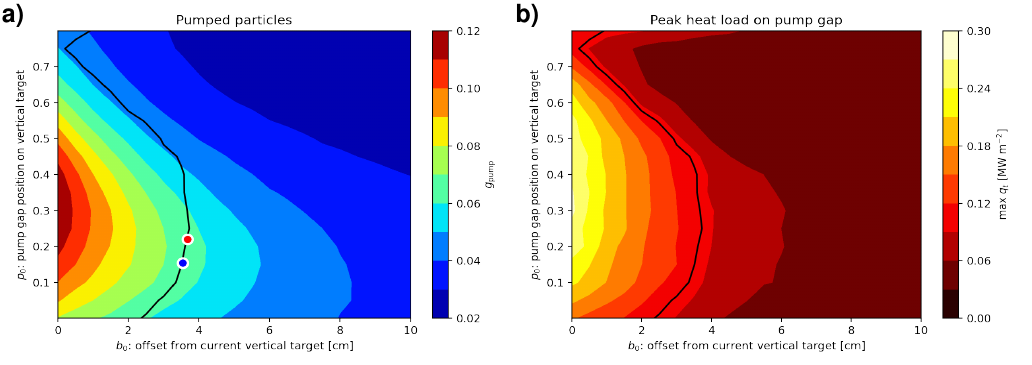}
\caption{Parameter scan of the pump gap position $p_0$ and vertical target offset $b_0$: (a) resulting pumped particle flux \gpump, and (b) peak heat load on pump gap $\max q_t|_{\mathrm{pump}} \, = \, \PSOL^\ast \, \cdot \, \qpeak|_{\mathrm{pump}}$ based on $\PSOL^\ast$ from the EMC3-EIRENE simulation. The (arbitrary) contour line $\qpumplimit \, = \, 0.1 \, \mega\watt \, \meter^{-2}$ is shown in black.
Colored dots mark the optimal configuration (maximal \gpump) with upstream conditions (red) and target averaged conditions (blue) for (\nedge, \Tedge). The latter is indistinguishable from the optimal configuration for (\nbg, \Tbg) for which \gpump values are shown here.}
\label{fig:VTA_pexhaust}
\end{center}
\end{figure}

Incident particles on the targets are either reflected or released as thermal molecules.
The current pump entrance is located to the side of the strike locations on the horizontal targets, but this may not be the best location to remove these particles.
In the following we will explore alternate positions of the pump entrance.
First, we replace the original gap with a baffle.
Then, we add a pump gap with fixed relative width $\wgap \, = \, 0.2$ along the vertical target at the relative position $p_0 \, \in \, [0, 1-\wgap]$ from the inside corner as highlighted in magenta in figure \ref{fig:shape_parameters}.
It should be noted that we can maximize \gpump by maximizing \wgap, but this would also increase the number of particles that can return from the sub-divertor.
By fixing \wgap (and \epump), we essentially assume equivalent sub-divertor / cryo-pump conditions (which we leave as a separate problem for another time).
We keep $\sigma \, = \, 10 \, \deg$ fixed and conduct a parameter scan for $(b_0, \, p_0)$, i.e. we push the vertical target back in order to avoid heat loads while scanning the pump gap position along the vertical direction.
Results are shown in figure \ref{fig:VTA_pexhaust}.
It can be seen that the best \gpump is achieved at $p_0 = 0.2 - 0.4$ without pushing back the vertical target.
However, this implies significant heat loads to the potential pump gap which must be avoided.
As shown earlier, this can be mitigated by introducing an offset $b_0$ for the vertical target position.
Nevertheless, this also implies reduced divertor closure which is detrimental for \gpump and \gfast.
As it turns out, optimizing for \gpump appears to be incompatible with minimizing $\qpeak|_{\mathrm{pump}}$, and one needs to find the right trade-off between the two.

This trade-off depends on the actual heat load the pump duct entrance behind the gap can tolerate.
Nevertheless, we only consider the heat load on the pump gap itself for convenience.
We take $\PSOL^\ast$ from the EMC3-EIRENE simulation in order to obtain absolute values $q_t$ from the proxy calculation $h_t$, and we set the arbitrary limit $\qpumplimit \, = \, 0.1 \, \mega\watt \, \meter^{2}$ (black contour lines in figure \ref{fig:VTA_pexhaust}).
Based on this constraint, we find the optimal configuration at $(b_0, \, p_0) \, = \, (3.5 \, \centi\meter, \, 0.15)$ which results in $\gpump \, = \, 0.065$.
There results are based on $(\nedge, \, \Tedge) \, = \, (\nbg, \, \Tbg)$, and it appears that the optimal configuration is rather robust with respect to this choice.
In fact, the optimal configuration is indistinguishable from the one found with target averaged conditions from the EMC3-EIRENE simulation (blue dot in figure \ref{fig:VTA_pexhaust}), albeit with a somewhat lower value of $\gpump \, = \, 0.052$ in the latter case.
This is consistent with figure \ref{fig:pexhaust_parameters} (b) which showed roughly similar \gpump values for these conditions.
For upstream conditions (red dot), on the other hand, the best achievable value is $\gpump \, = \, 0.029$, which is considerably smaller.
Nevertheless, the optimal configuration is found at $(b_0, \, p_0) \, = \, (3.7 \, \centi\meter, \, 0.22)$, which is not too far off from the other cases.
This suggests that there is some resiliency in the geometry optimization with respect to uncertainties in plasma conditions.

\subsection{Multivariate optimization}

\begin{figure}
\begin{center}
\includegraphics[width=160mm]{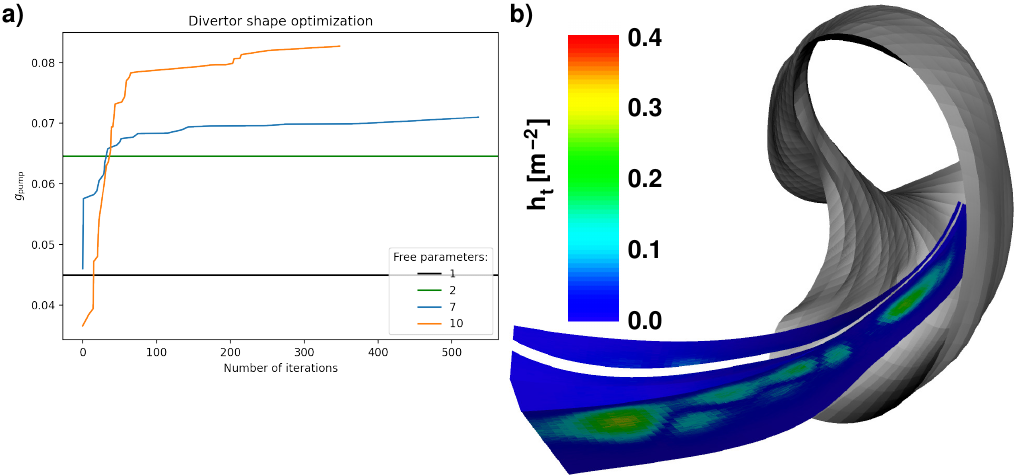}
\caption{(a) Optimization progress for two scenarios:
1) where the vertical target itself remains fixed with offset parameters $b_0 \, = \, 2 \, \centi\meter$, $\sigma \, = \, 10 \, \deg$, $\varphi_0 \, = \, -9.5 \, \deg$ (blue), and
2) where $b_0$, $\sigma$ and $\varphi_0$ are additional shape coefficients to be optimized (orange).
Reference values for maximal \gpump are taken from the constrained region of figure \ref{fig:VTA_pexhaust} (green) and its subset at fixed $b_0 \, = \, 2 \, \centi\meter$ (black).
(b) Optimal pump gap configuration with vertical target adjustments $b_0 \, = \, 4.2 \, \centi\meter$, $\sigma \, = \, 4.8 \, \deg$, $\varphi_0 \, = \, -16.3 \, \deg$.
}
\label{fig:optimization}
\end{center}
\end{figure}

As a final example, we will now combine the previous observations into a multivariate optimization problem.
As for the initial fit of the divertor geometry, we now consider the case where the pump gap position $p_0$ is a function of $\varphi$ and use again a cubic spline representation with 7 equidistant knots and shape coefficients.
The optimization problem is

\begin{equation}
\begin{aligned}
\underset{\vec{c}}{\mathrm{minimize}} \quad & -\gpump(\vec{c}) \\
\mathrm{subject \, to} \quad & \PSOL^\ast \, \cdot \, \qpeak|_{\mathrm{pump}} \, \le \, 0.1 \, \mega\watt \, \meter^{-2}
\end{aligned}
\end{equation}

\noindent where $\vec{c}$ are the shape coefficients.
A number of different methods exist to solve this kind of problem, and a review of those is out of scope here.
Here we take the particle based approach of the proxy calculations to the next level and implement a particle swarm optimization (PSO) method (see \ref{sec:PSO} for a more detailed description of this method).
Essentially, PSO is a metaheuristic which tries to iteratively improve a population of candidate solutions (particles).
Advantages are that it is relatively easy to implement and that it does not require the computation of gradients unlike many classical methods.
This is of particular relevance because of the intrinsic noise associated with the particle based approach in the underlying objective function.

Figure \ref{fig:optimization} (a) shows the optimization progress for two scenarios:
1) where the vertical target itself remains fixed with offset parameters $b_0 \, = \, 2 \, \centi\meter$, $\sigma \, = \, 10 \, \deg$, $\varphi_0 \, = \, -9.5 \, \deg$ (blue), and
2) where $b_0$, $\sigma$ and $\varphi_0$ are additional shape coefficients to be optimized (orange).
Reference values for maximal \gpump are taken from the constrained region of figure \ref{fig:VTA_pexhaust}.
For the first scenario, we consider the univariate ($p_0$) optimization at fixed $b_0 \, = \, 2 \, \centi\meter$ (black) which shows that \gpump can be improved from 0.045 to 0.071 when the pump gap position is allowed to vary in toroidal direction.
For the second scenario where we include adjustments of the vertical target in the optimization, we find that we can further improve \gpump from the value of 0.065 of the $(p_0, b_0$) parameter scan to 0.083.
The resulting configuration is shown in figure \ref{fig:optimization} (b).
By including $\varphi_0$ in the optimization, we find that the required adjustment for the heat load constraint is a moderate offset of $b_0 \, = \, 4.2 \, \centi\meter$ that is introduced near the lower toroidal end of the target at $\varphi_0 \, = \, -16.3 \, \deg$ over a range of $\sigma \, = \, 4.8 \, \deg$.
Further improvements with less adjustment may be achieved by allowing for a variable gap width which can be narrower where needed without sacrificing exhaust efficiency elsewhere.
But this remains future work.


\section{Conclusions}

The FIREFLY package for rapid evaluation of divertor designs has been introduced.
It leverages field line reconstructions from a magnetic flux tube mesh for fast approximation of divertor heat loads.
It is found that the heat load distribution from an EMC3-EIRENE simulation for W7-X can be reproduced with careful choice of model parameters.
In particular, it is found that the density parameter for the simplified calculation needs to be smaller than characteristic densities in the EMC3-EIRENE simulation.
This is likely related to the missing heat convection and simplified heat conduction and boundary condition in FIRELY and EMC3-Lite.
Follow-up studies will show how the best fitting model parameters depend on power settings and magnetic field configurations.

The heat load distribution from the simplified heat transport calculation is then used to sample neutral particles.
These are followed until they are either ionized or removed at a pumping surface.
The EIRENE code is leveraged for surface reflections as well as atomic and molecular reactions, however, particles are tracked in a constant background plasma.
This can be used to approximate the efficiency of particle removal through pumping gaps.
A parametrization of the divertor geometry in W7-X has been introduced as the foundation for further geometry adjustments.
Initial parameter scans have shown the potential for improved particle exhaust by introducing a small offset to the vertical target along with moving the pump gap away from the corner with the horizontal target.
Finally, we have integrated the proxy calculations into a multivariate optimization procedure to maximize particle removal while avoiding heat loads to the potential pump gap.


\appendix
\section*{Acknowledgments}
This work was supported by the U.S. Department of Energy under awards No. P-240001537 and DE-SC0014210.

\section{Computation times}

\begin{figure}
\begin{center}
\includegraphics[width=160mm]{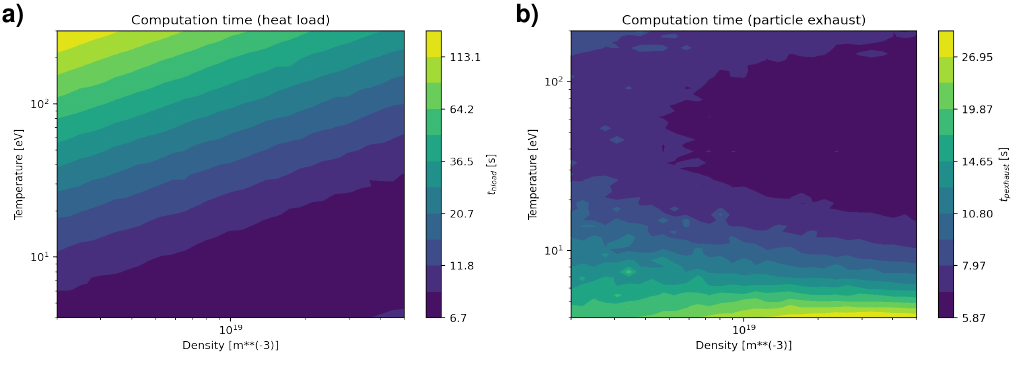}
\caption{Computation times for the parameters scans in figure \ref{fig:hload_parameters} and \ref{fig:pexhaust_parameters} which have been conducted on NERSC's Perlmutter system. Simulations are conducted with $10^6$ particles split over 2 nodes with 128 processes each.}
\label{fig:computation_time}
\end{center}
\end{figure}

The simplified heat load and particle exhaust calculations have been carried out on the Permutter system at NERSC.
Computation times for the parameters scans in figure \ref{fig:hload_parameters} and \ref{fig:pexhaust_parameters} are shown in figure \ref{fig:computation_time}.
For the heat load proxy, computations are fastest at low temperature \Tbg and high density \nbg which implies relatively large cross-field transport compared to transport along field lines.
For the particle exhaust proxy, on the other hand, charge exchange happens more frequently with respect to ionization under these conditions which implies that computations are slower.
Here, all heat load simulations are conducted with the same time step $\tau \, = \, 5 \cdot 10^{-7} \, \second$ - even though the fastest simulations in figure \ref{fig:computation_time} (a) violate the small $\delta_\parallel / \lambda_{T \parallel}$ requirement.
Nevertheless, the same parameter scan with fixed \ploss does not show any relevant differences to figure \ref{fig:hload_parameters}.

\section{Particle swarm optimization} \label{sec:PSO}

Particle swarm optimization (PSO) is a stochastic optimization method which is inspired by the social behavior of bird flocking or fish schooling \cite{Kennedy1995, Eberhart1995}.
In PSO, a population of candidate solutions (referred to as particles) are iteratively improved by moving them around through search space.
Each particle $i$ has the following parameters:

\begin{center}\begin{tabular}{ll}
$\vec{x}_i$:	& the {\it current position} of the particle \\
$\vec{v}_i$:	& the {\it current velocity} of the particle \\
$\vec{p}_i$:	& the {\it personal best known position} of the particle
\end{tabular}\end{center}

The personal best known position $\vec{p}_i$ is either the current position $\vec{x}_i$ or one of its previous values - whichever yields the highest fitness value for that particle.
For minimization of a multi-variate function $f(\vec{x})$, higher fitness is determined by a lower function value.
The movement is influenced by each particles best known position $\vec{p}_i$, but also guided towards the {\it global best known position} $\vec{g}$ (variations of the algorithm exist where a local neighborhood best $\vec{L}_i$ is used instead of $\vec{g}$).
For the next iteration ($t + 1$) of the swarm, the velocity of particle $i$ is updated as follows:

\begin{equation}
\vec{v}_i(t+1) \, = \, w \, \vec{v}_i(t) \, + \, c_1 \, \vec{r}_1 \odot (\vec{p}_i \, - \, \vec{x}_i(t)) \, + \, c_2 \, \vec{r}_2 \odot (\vec{g} \, - \, \vec{x}_i(t)) \label{eq:PSO-Vupdate}
\end{equation}

\noindent where $w$ is referred to as the {\it inertia weight} while $c_1$ and $c_2$ are often called {\it cognitive coefficient} and {\it social coefficient}, respectively \cite{Shi1998}.
One particular combination is $w = 0.729$ and $c_1 = c_2 = 1.494$ \cite{Eberhart2000}, but other combinations are possible (within certain limits for stability \cite{Clerc2002, Trelea2003}).
The second and third terms include uniform random weights $\vec{r}_1$ and $\vec{r}_2$ which are applied element wise ($\odot$).
The position update is then

\begin{equation}
\vec{x}_i(t+1) \, = \, \vec{x}_i(t) \, + \, \vec{v}_i(t+1) \label{eq:PSO-Xupdate}
\end{equation}

\noindent where $\vec{p}_i$ and $\vec{g}$ are updated when applicable (not every iteration results in an improvement).
Particles are initialized with random locations and velocity.
Velocities are clamped to $\pm \vmax$ with $\vmax \, = \, \vec{b}_u \, - \, \vec{b}_l$ throughout the procedure, where $\vec{b}_u$ and $\vec{b}_l$ are the upper and lower bounds of the search space, respectively.
Particles are allowed to {\it fly} beyond those boundaries, but $f(\vec{x})$ is not evaluated there.
Likewise, $\vec{p}_i$ and $\vec{g}$ are not updated here when given constraints are not met.
More sophisticated methods for constraints \cite{Elsayed2013, Kohler2019} can be implemented later.


\section*{References}


\end{document}